\documentclass[prd,nofootinbib,tightenlines,showpacs]{revtex4}
\def\journal#1, #2, #3#4, #5#6#7#8    {
    {#1~} {#2}  (#5#6#7#8) #3#4}

\def\plb{\journal Phys. Lett. B, }

\newcommand{\beq}[1]{\begin{equation}\label{#1}}
\newcommand\eeq{\end{equation}}
\newcommand{\ba}[1]{\begin{eqnarray}\label{#1}}
\newcommand{\baa}{\begin{eqnarray}}
\newcommand\ea{\end{eqnarray}}
\newcommand{\bee}{\begin{equation}}
\newcommand{\br}[1]{\overline #1}
\def\nn{\nonumber \\}
\def\l{\lambda}

\newcommand{\ho}{harmonic oscillator}
\newcommand{\h}{Hamiltonian}

\newcommand{\B}[1]{{\bf #1}}
\def\hlf{\frac{1}{2}}

\newcommand{\lsim}{\stackrel{\rm <} {\scriptstyle \sim}}

\begin{document}
\title{Harmonic oscillator with minimal length 
uncertainty relations and ladder operators}
\author{Ivan Dadi\'c}
\email{dadic@thphys.irb.hr}
\author{Larisa Jonke}
\email{larisa@irb.hr}
\author{Stjepan Meljanac}
\email{meljanac@irb.hr} 
\affiliation{Theoretical Physics Division,\\
Rudjer Bo\v skovi\'c Institute, P.O. Box 180,\\
HR-10002 Zagreb, Croatia}

\begin{abstract}
We construct creation and annihilation operators for deformed \ho s with 
minimal length uncertainty relations. We discuss a possible generalization
to a large class of deformations of cannonical commutation relations.
We also discuss dynamical symmetry of noncommutative harmonic oscillator.
\end{abstract}

\vspace{1cm}
\maketitle

\section{Introduction}

The existence of the minimal length, at least at the Planck scale, 
seems to  be a general feature of any quantum theory of gravity. 
Test particles of sufficiently high energy
for probing small scales curve gravitationally,  and thereby, disturb the 
very space-time they are probing. 
Both, perturbative string theory considerations \cite{Gross:1987ar}
 and black hole 
physics \cite{mar}  give rise to modified space-momentum uncertainty 
relations\footnote{See Ref.\cite{yon} for space-time
uncerainty relations, originating from nonperturbative string theory.}
that imply the existence of a minimal length.
The investigation of cosmological consequences 
of these modified space-momentum uncertainty 
relations has been intensified lately. It appears that minimal length 
uncertainty relations (MLUR) can offer some answers to the 
problem of black hole remnants \cite{Adler:2001vs}, 
the trans-Planckian problem of 
inflation \cite{sloth}, and the cosmological constant problem \cite{minic}. 
On the other hand, one can discuss MLUR in the context of deformation of 
quantum mechanics, since the uncertainty relations and the underlying canonical 
commutation relations are at the heart of  quantum mechanics. 
The generalized quantum-theoretical framework which 
implements the appearence of MLUR was discussed
in Ref.\cite{Kempf:1994su}, and 
the formalism obtained was applied to the \ho \ case.
Recently, the 
exact solution of the \ho \ in arbitrary dimensions with MLUR has been
found \cite{Chang}.

In one spatial dimension, the generalized commutation relation can be written
as
$$[x,p]=i f(x,p),$$
with $x$ and $p$ hermitian operators and
$[f(x,p)]^{\dagger}=f(x,p)$
Then, in any physical state one finds
$$\Delta x\Delta p\geq \hlf|\overline{ f(x,p)}|,$$
where we define for any operator $O$
$${\br O} =\langle\Psi| O| \Psi\rangle,\; 
\Delta O=\sqrt{\langle\Psi|(O-\br O)^2|\Psi\rangle}.$$
The operator function $f(x,p)$ can be treated as a smooth deformation
of ordinary 
quantum mechanics with $f_0(x,p)=\hbar$.
Note that 
the
limit to  classical mechanics, $f(x,p)\rightarrow 0$, is not smooth.
In this paper we  restrict ourselves to smooth deformations of 
quantum mechanics, only. 

A number of physical
problems can be expressed as a deformed 
\ho \ with generalized commutation relations,
such as a singular
Calogero potential in one dimension \cite{poly}, the Landau
problem in two dimensions, the \ho \ in  noncommutative plane 
(see Ref.\cite{nch} and references therein).
Therefore, it is interesting to analyze 
the harmonic oscillator case  
with a general 
deformation $f(x,p)$.
For a large class of smooth deformations  $f(x,p)$, one expects a 
smooth deformation of a ground state (Gaussian function) and a  smooth 
deformation of excited states (Hermite polynomials).

For smooth deformations,
one also expects that the corresponding Fock space and 
the creation and annihilation operators can be smoothly deformed. Of course, 
an important question is to find a general method for constructing ladder 
operators for any smooth deformation  $f(x,p)$.
Even when such deformed oscillators can be solved exactly, there is no
well-defined method for constructing  creation and annihilation 
operators for the corresponding Fock space. 
The "second quantization" is 
crucial in the analysis of many-body problems
 and in discussing the dynamical
symmetry algebra of the underlying problem.

In this brief report we concentrate on the special class of deformations
$f(x,p)=1+\beta p^2, \beta\geq 0$ and construct ladder operators for 
the corresponding \ho \ problem. We also consider the $D$-dimensional case with 
$SO(D)$ rotational invariance. These deformations are physically motivated by 
generalized uncertainty relations implying the minimal length $\Delta x\geq l_{\rm min}$
and possess the  interesting UV/IR connection.
This is a simple quantum-mechanical example inspired by string theory 
and cosmology. 

\section{Harmonic oscillator in one dimension}

The position and momentum operators obeying ($\hbar =1$)
\bee\label{def1}
[X,P]=i(1+\beta P^2),\eeq
are represented in  momentum space by
$$X=i\left[(1+\beta p^2)\frac{\partial}{\partial p}+\beta p\right],\;{\rm and}\;P=p.$$
The Schr\"odinger equation in  momentum space corresponding to the 
\ho\footnote{It is not quite appropriate to call \h \ (\ref{h2}) 
a harmonic oscillator, 
since its equations of motion, with the commutation relation (\ref{def1}), 
are not harmonic.  It 
would be more fitting to call it a quadratic oscillator.
However, in this paper we use the same terminology as in 
Refs.\cite{Kempf:1994su,Chang}, i.e., (deformed) harmonic oscillator.}
with the \h \ $(\omega=m=1)$
\bee\label{h2}
H=\hlf(P^2+X^2)= \hlf\left[-\left((1+\beta p^2)\frac{\partial}{\partial p}
          \right)^2
        - 2\beta p 
          \left( (1+\beta p^2)\frac{\partial}{\partial p}
          \right)
        - 2\beta^2 p^2
        - \beta
     +p^2\right]
\eeq
leads  to the eigenvalue problem $H\Psi_n=E_n\Psi_n$. The exact 
solutions \cite{Chang} are
\baa\label{sol}
E_n&=&(n+\hlf)\sqrt{1+\frac{\beta^2}{4}}+\left(n^2+n+\hlf\right)
\frac{\beta}{2},\nn
\Psi_n&=&2^{\l}\Gamma(\l)\sqrt{\frac{n!(n+\l)\sqrt\beta}{2\pi\Gamma(n+2\l)}}
c^{\l+1}C^{\l}_n(s),\ea
where $n=0,1,2,\ldots,$ and $2\l=1+\sqrt{1+4/\beta^2}$,
\ba 1
c&=&\frac{1}{\sqrt{1+\beta p^2}},\nn
s&=&\frac{\sqrt{\beta}p}{\sqrt{1+\beta p^2}},\nonumber\ea
and $C^{\l}_n(s) $ is the Gegenbauer polynomial:
\beq G
C^{\l}_n(s)=\frac{(-)^n}{2^nn!}\frac{\Gamma(2\l+n)\Gamma(\frac{2\l+1}{2})}
{\Gamma(2\l)\Gamma(\frac{2\l+1}{2}+n)}(1-s^2)^{1/2-\l}\frac{d^n}{ds^n}
(1-s^2)^{\l+n-1/2},\eeq
satisfying the recursive relations \cite{abr}
$$(n+1)C^{\l}_{n+1}(s)=\left[(2\l+n)s-c^2\frac{d}{ds}\right]C^{\l}_n(s).$$
For the normalized functions $\Psi_n(s)$, the recursive relations are
\bee\label{p1}
(n+1)\Psi_{n+1}^{\l}(s)=(n+\l-1)\frac{{\cal N}_{n+1}}{{\cal N}_{n}}
s\Psi_{n}^{\l}(s)
-\frac{{\cal N}_{n+1}}{{\cal N}_{n}}c^2\frac{d}{ds}\Psi_{n}^{\l}(s),\eeq
where 
$${\cal N}_n=2^{\l}\Gamma(\l)\sqrt{\frac{n!(n+\l)\sqrt\beta}{2\pi\Gamma(n+2\l)}}.$$

Let as define
$$\Psi_n(s)=\frac{b^{\dagger n}}{\sqrt{n!}}|0\rangle,\; n=0,1,2,\ldots,$$
where $b^{\dagger}$ and $b$ are bosonic operators, $[b,b^{\dagger}]=1$,
with the number operator $N=b^{\dagger}b$ and
$$[N,b^{\dagger}]=b^{\dagger},\;[N,b]=-b.$$
Now we easily find
\ba b
&&b^{\dagger}\Psi_n(s)=\sqrt{n+1}\Psi_{n+1}(s),\nn
&&b\Psi_n(s)=\sqrt{n}\Psi_{n-1}(s),\nn
&&N\Psi_n(s)=n\Psi_n(s).\nonumber\ea
Using the recursive relations (\ref{p1}) we obtain
\ba a
&&b^{\dagger}=\left[s(N+\l-1)-c^2\frac{d}{ds}\right]\sqrt{\frac{N+\l+1}{(N+\l)(N+2\l)}},\nn
&&b=\sqrt{\frac{N+\l+1}{(N+\l)(N+2\l)}}\left[(N+\l)s+c^2\frac{d}{ds}\right].
\ea
From the \h \ 
$H=\beta\l\left(N+\hlf\right)+\hlf\beta N^2$ we can express the number operator
$$N=\frac{1}{\beta}\left[-\beta\l+\sqrt{\beta^2\l^2+\beta(2H-\beta\l)}\right].$$
In the limit $\beta\rightarrow 0,\;\beta\l\rightarrow 1$, the  \h \ becomes
$H=N+1/2$ and the wave functions and the ladder operators smoothly 
go to the ordinary \ho \ case:
\ba l
&&\lim_{\beta\to 0}c^{\l+1}=\exp{(-p^2/2)},\nn
&&\lim_{\beta\to 0}\Psi_n(s)={\cal N}_n(0)\exp{(-p^2/2)} H_n(p),\nn
&&\lim_{\beta\to 0}b^{\dagger}=\frac{1}{\sqrt{2}}\left(p-\frac{d}{dp}\right),\nn
&&\lim_{\beta\to 0}b=\frac{1}{\sqrt{2}}\left(p+\frac{d}{dp}\right).\nonumber\ea

We can obtain an interesting result if we define operators 
$A,A^{\dagger}$  
\bee\label{ppp}
A^{\dagger}=b^{\dagger}\sqrt{\left(1+\frac{N}{2\l}\right)\beta \l},\;
A=\sqrt{\left(1+\frac{N}{2\l}\right)\beta \l}\;b,\eeq
with the commutator
$[A,A^{\dagger}]=\beta\l+\beta N$. Then we can write the \h \ 
as
$$H
=\beta\l\left(N+\hlf\right)+\hlf\beta N^2=\hlf\{A,A^{\dagger}\}.$$
The deformed oscillators (\ref{ppp}) are examples 
of the general deformed oscillator mapping, see Refs.\cite{po}.
Furthermore, we   redefine the operators $A,A^{\dagger}$ for $\beta >0$:
$$J_-=\sqrt{\frac{2}{\beta}}A,\;J_+=\sqrt{\frac{2}{\beta}}A^{\dagger},\; J_0=N+\l.$$
In this way  they become generators of $SU(1,1)$ algebra:
$$[J_-,J_+]=2J_0,\;[J_0,J_{\pm}]=\pm J_{\pm}.$$
The deformed \ho \ in one dimension, Eqs.(\ref{def1}) and (\ref{h2}), 
possesses a hidden $SU(1,1)$ symmetry for 
$\mbox{$\beta >0$}$.
The same hidden symmetry was found in the quantum system with an infinitely 
deep square-well potential \cite{Ma}.
For $\beta <0$, the algebra of the 
operators $A,A^{\dagger}$ has a finite dimensional 
representation if $2\l$ is an integer. 
In this case, there is no minimal length and the 
system becomes parafermionic. It corresponds to a hidden $SU(2)$ symmetry.

The benefit of our construction of ladder operators is obvious when considering 
the many-body 
problem. The simplest way to consider $N$ free deformed \ho s is to define
$$H=\hlf\sum_{i=1}^N\{A_i,A_i^{\dagger}\},$$
with the algebra of multimode oscillators \cite{multi}
$$[A_i,A_j^{\dagger}]=(\beta\l+\frac{\beta}{2}N)\delta_{i,j},\;[A_i,A_j]=[A_i^{\dagger},
A_j^{\dagger}]=0.$$
The procedure for finding the algebra of observables and dynamical symmetry 
algebra is 
the same as presented in Refs.\cite{op}.

A large class of smooth deformations of the one-dimensional \ho \ case can be 
described by the  wave function $\Psi_n=\Psi_0(s)P_n(s)$, where $s=s(p)$ is 
an arbitrary function of momentum,  
$\Psi_0(s)$ is a smooth deformation of the Gausssian, and the orthogonal 
polynomial $P_n(s)$ is one of the following three types, up to simple 
deformations:
\baa\label{ortho}
&& P^{(1)}_n(s)\propto(as+b)^{-\alpha}(cs +d)^{-\beta}\frac{d^n}{ds^n}\left[
(as+b)^{\alpha+n}(cs +d)^{\beta+n}\right]\;,\nn
&& P^{(2)}_n(s)\propto(as+b)^{-\alpha}e^{+\beta s}\frac{d^n}{ds^n}\left[
(as+b)^{\alpha+n}e^{-\beta s} \right]\;,\nn
&& P^{(3)}_n(s)\propto e^{+(\alpha s^2+\beta s)}\left[as+b\frac{d}{ds}\right]^n
e^{-(\alpha s^2+\beta s)}\;.\ea
For example, for the two-body Calogero model the \h \ for relative motion is 
$H=\hlf[p^2+x^2+\nu(\nu-1)/x^2]$ and the polynomial part of the 
wave function is 
$$P^{(3)}_n(x)\propto e^{\hlf x^2} \left[x-\frac{d}{dx}+\frac{\nu}{x}(1-K)\right]
^ne^{-\hlf x^2},$$ where $K$ is exchange operator, $Kx=-xK$, and  
on symmetric states the $\nu$ part is zero.

Using recursive relations for the orthogonal polynomials (\ref{ortho})
 one can construct 
creation and annihilation operators for a large class of deformations, 
simply following the procedure outlined in this section.

\section{Harmonic oscillator in D-dimensions}

In more than one dimensions, the modified commutation relation can be generalized to the 
tensorial form:
\ba 3
&&[X_i,P_j]=i(\delta_{ij}+\beta P^2\delta_{ij}+\beta'P_iP_j),\nn
&&[P_i,P_j]=0,\; X_i^{\dagger}=X_i,\; P_i^{\dagger}=P_i.\ea
Then, the commutation relations among the coordinates $X_i$ are almost uniquely determined
by the Jacobi identity (up to possible extensions, see Kempf \cite{Kempf:1994su}).
The operators $X_i$ and $P_j$ satisfying the relations (\ref{3}) are realized in momentum 
space as
\ba 4
&&X_i=i\left[(1+\beta p^2)\frac{\partial}{\partial p_i}+\beta'p_ip_j\frac{\partial}
{\partial p_j}+\left(\beta+\frac{D+1}{2}\beta '\right)p_i\right],\nn
&&P_i=p_i. \ea
The condition for  the existence of  minimal length  is $l^2_{\rm min}=D\beta
+\beta'>0$.
The \h \ for $D-$dimensional deformed \ho 
\beq z 
H=\hlf\left(\B P^2+\B X^2\right)=\hlf\sum_{i=1}^D\left(P^2_i+X^2_i\right)\eeq
possesses $O(2D)$ rotational symmetry in $2D$-phase space.
However, the transformation Eq.(\ref{4}) and hence all commutation relations
preserve the same form under $O(D)$ transformations: $X_i'=R_{ij}X_j,P_i'=R_{ij}P_j$
and $x_i'=R_{ij}x_j,p_i'=R_{ij}p_j$, where $R\in O(D)$. 
Hence, the dynamical symmetry 
of the problem at hand is $O(D)$.
Therefore we   assume that the 
energy eigenstates in  the momentum space when expressed in terms of radial 
momenta can be written as
a product of spherical harmonics and a radial wave function:
 $$\Psi_D(\B p)=Y_{l_{D-1}\cdots l_2l_1}(\Omega)R(p),\;p=\sqrt{\B p^2},\;
 l=l_{D-1}\geq \cdots l_2\geq |l_1|.$$
Then one can perform the replacement
\begin{eqnarray}
\sum_{i=1}^N \frac{\partial^2}{\partial p_i^2}
& = & \frac{\partial^2}{\partial p^2}
    + \frac{D-1}{p}\frac{\partial}{\partial p} 
    - \frac{L^2}{p^2}\;, \cr
\sum_{i=1}^N p_i\frac{\partial}{\partial p_i} 
& = & p\frac{\partial}{\partial p}\;,
\end{eqnarray}
where 
\begin{equation}
L^2 = l (l + D -2 )\;,\qquad
l = 0,1,2,\ldots\;.
\end{equation}
For example, in two-dimensional case, $  Y_m(\phi)=\exp{(-im\phi)}/\sqrt{2\pi}$
and $l=|m|,\;m\in Z$.
Note that 
$$\left(\frac{\partial}{\partial p}\right)^{\dagger}=-\frac{\partial}
{\partial p}-
\frac{D-1}{p}.$$

We therefore find  that
the Schr\"odinger equation for the $D$-dimensional 
oscillator can be reduced to the 1-dimensional problem 
for the radial wave function $R(p)$.
The energy eigenvalues for $\beta+\beta'>0$ are given by
\begin{eqnarray}\label{enl}
E_{nl}
& = &
\left[
  \left( n + \frac{D}{2} \right) 
        \sqrt{ 1 + \left\{ \beta^2 L^2 
                         + \frac{ (D\beta + \beta')^2 }{ 4 }
                   \right\}
             }
\right. \cr
& & \qquad\left.
+ \hlf\left\{ (\beta + \beta')\left( n + \frac{D}{2}\right)^2
        + (\beta - \beta')\left(L^2 + \frac{D^2}{4}\right)
        + \beta'\frac{D}{2}
  \right\}
\right]\;,
\end{eqnarray}
where $n=2n'+l$ and $n'$ and $l$ are non-negative integers. The D=1 case 
can be reproduced by setting $L^2=0$ and $\beta'=0$. The states with $l=0$ are 
even eigenstates ($n=2n'$) and states with $l=1$ are  odd eigenstates ($n=2n'+1$).
The normalized energy eigenfunctions are
\begin{equation}
R_{n\ell}(p) = 
\sqrt{ \frac{ 2 (2n'+a+b+1)\, n'!\, \Gamma(n'+a+b+1) }
             { \Gamma(n'+a+1) \Gamma(n'+b+1) }
     }\,
(\beta+\beta')^{D/4}\, c^{\lambda+\delta}\, s^\ell\, P_{n'}^{(a,b)}(z)\;,
\end{equation}
where $P_n^{(a,b)}(z)$ is the Jacobi polynomial,
\begin{eqnarray}
c & = & 
  \frac{1}{ \sqrt{ 1 + (\beta+\beta')p^2 } }\;, \;\;
s  =  
  \frac{ \sqrt{\beta + \beta'}\,p }{ \sqrt{1 + (\beta+\beta')p^2} }\;,\cr
z & = & 2s^2 -1,\;\;
\delta = \frac{\beta+\beta'(D+1)/2}{\beta+\beta'},\cr
a & = & \l-\frac{1+\beta(D-1)/(\beta+\beta')}{2},\;\;
b  =  \frac{D}{2}+l-1,\;
\end{eqnarray}
and $\l$ is the positive root of the equation
$$\lambda^2 - \lambda \left[ 1+(D-1)\frac{\beta}{\beta+\beta'} \right]
- \left(\frac{\beta}{\beta+\beta'}\right)^2 L^2 - \frac{1}{(\beta+\beta')^2}
= 0\;.$$
Interesting cases are i)$\beta=0,\beta'>0$, ii)$\beta+\beta'=0,\beta>0$, 
and iii)$2\beta=\beta'>0$, 
see Refs.\cite{Kempf:1994su,brau}.
Using recursive relations for  Jacobi polynomials \cite{abr}
\baa\label{jprr}
&&(2n+a+b)(1-x^2)\frac{d}{dx}P_n^{(a,b)}(x)=n[(a-b)-(2n+a+b)x]P_n^{(a,b)}(x)+
2(n+a)(n+b)P_{n-1}^{(a,b)}(x),\nn
&&2(n+1)(n+a+b+1)(2n+a+b)P_{n+1}^{(a,b)}(x)=\nn
&&\hspace{1.3cm}(2n+a+b+1)[(2n+a+b)(2n+a+b+2)x+a^2-b^2]
P_n^{(a,b)}(x)\nn 
&&\hspace{4.9cm}-2(n+a)(n+b)(2n+a+b+2)P_{n-1}^{(a,b)}(x),\ea
we find recursive relations for the energy eigenfunctions $\Psi_{nl}$ and hence we 
find the unique ladder (creation and annihilation) operators for the radial excitations
$a^{\dagger}_r(\beta,\beta',D,l)$ and $a_r(\beta,\beta',D,l)$. Demanding
$ a_r|0,l\rangle=0$ and 
\bee\label{fock}
\Psi_{nl}=\frac{(a_r^{\dagger})^{n'}}{\sqrt{n'!}}Y_{l_{D-1}\cdots l_1,m}=
\frac{(a_r^{\dagger})^{n'}}{\sqrt{n'!}}|0,l\rangle,\;
n=2n'+l,\;n',l=0,1,2,\ldots,\eeq
and 
\ba x
a_r^{\dagger}\Psi_{nl}=\sqrt{n'+1}\Psi_{n+2,l},\;
a_r\Psi_{nl}=\sqrt{n'}\Psi_{n-2,l},\nn
a_r^{\dagger}a_r=N',\;a_ra_r^{\dagger}=N'+1,\nonumber\ea
we find
\baa\label{ar1}
&&a^{\dagger}_r(\beta,\beta',D,l)=\left[C'_{1N'}+xC'_{2N'}-(1-x^2)
\frac{d}{dx}(2N'+a+b)(2N'+a+b+2)\right]\times\nn
&&\times \frac{1}{2(2N'+a+b)}\left[\frac{(2N'+a+b+3)}{(2N'+a+b+1)(N'+a+b+1)
(N'+a+1)(N'+b+1)}\right]^{1/2},\nn
&&a_r(\beta,\beta',D,l)=\left[C_{1N'}+xC_{2N'}+(1-x^2)
\frac{d}{dx}(2N'+a+b)\right]\times\nn
&&\times \frac{1}{2}\left[\frac{(2N'+a+b-1)}{(2N'+a+b+1)(N'+a+b)
(N'+a)(N'+b)}\right]^{1/2},\ea
where $C'_{1N'}, C'_{2N'}, C_{1N'}, C_{2N'}$ are given by the following 
expressions:
\baa\label{cN}
C'_{1N'}&=&(a-b)[(2N'+a+b+1)(N'+a+b)+N']\nn
&-&(2N'+a+b)(2N'+a+b+2)\frac{\l+1-l}{2}\;,\nn
C'_{2N'}&=&(2N'+a+b)(2N'+a+b+2)\left(N'+a+b+1-\frac{\l+1+l}{2}\right),\nn
C_{1N'}&=&(2N'+a+b)\frac{\l+1-l}{2}-N'(a-b),\nn
C_{2N'}&=&(2N'+a+b)\left(N'+\frac{\l+1+l}{2}\right).\nonumber\ea
The states $|0,l>$ with energy $E_{l,l}$, Eq.(\ref{enl}),
are $\left[{{D+l-1}\choose{D-1}}-
{{D+l-3}\choose{D-1}}
\right]$-fold degenerate, and can be represented by an irreducible 
representation of group $SO(D)$.
If $\beta=\beta'=0$, the degeneracy  of states with energy 
$E_n=n+D/2$ is larger, i.e., ${N+l-1}\choose{D-1}$,
corresponding to the totally symmetric irreducible representation of 
$SU(D)$ dynamical symmetry.
Note that one can simply generalize the transformations (\ref{4}), i.e., 
commutation relations Eq.(\ref{3})
by taking $\beta\delta_{ij}\rightarrow\beta_{ij},\;
\beta' P_iP_j\rightarrow\beta_{ij}'P_iP_j$ to 
obtain different dynamical symmetries of the type $\prod O(D_i)$, 
$\sum D_i=D$.

In the limit $\beta\rightarrow 0$, $\beta'\rightarrow 0$ and 
$0\leq\beta/(\beta+\beta')\leq 1$, we find
\baa\label{alim}
a_r&=&\left[-2N'-l+p^2+p\frac{d}{dp}\right]\frac{1}{2\sqrt{N'+l-1+D/2}}\;,\nn
a_r^{\dagger}&=&\left[-2N'-l-D+p^2-p\frac{d}{dp}\right]
\frac{1}{2\sqrt{N'+l+D/2}}\;.\ea

For $D=1$ and $l=0$,
\baa\label{D1}
a_0&=&\frac{1}{2\sqrt{N'+1/2}}\left[p^2+p\frac{d}{dp}-2N'\right]\;,\nn
a_0^{\dagger}&=&\left[p^2-p\frac{d}{dp}-2N'-1\right]\frac{1}{2\sqrt{N'+1/2}}\;.
\ea

Let us briefly discuss the symmetry aspects of the deformed \ho , i.e., 
\ho \ defined on noncommutative space. The \h \ (\ref{z}) is invariant under
$O(2D)$ transformation in phase space. However, the symmetry of commutation 
relations may be quite different for different noncommutative spaces. 
In the case of ordinary cannonical variables
$x_i,p_i$, the symmetry is $Sp(2D)$, so that the dynamical symmetry of ordinary 
\ho \ in $D$ dimensions is $O(2D)\cap Sp(2D)=U(D)$.
A complete analysis  of phase space symmetry structure was performed
in Ref.\cite{nch} for noncommutative space in which the commutators of phase 
space variables are constants (c-numbers).
In the case of generalized commutation relations implying minimal length 
uncertainty relations, 
Eqs.(\ref{3}) and (\ref{4}), 
the commutators are  no longer c-numbers. However, they are 
invariant under $O(D)$ transformation,  $X_i'=R_{ij}X_j,P_i'=R_{ij}P_j$. 
Hence, the dynamical symmetry is $O(2D)\cap O(D)=O(D)$. Note that the 
symmetry of the 
transformed \h \ in terms of cannonical variables 
$H(X_i(x_i,p_i),P_i=p_i)$, 
is $O(D)$ and the symmetry preserving
cannonical commutation relations is $Sp(2D)$, so, again, the dynamical 
symmetry for the transformed system is $O(D)$. The system defined by 
cannonical commutation relations and the \h \ $H(X_i(x_i,p_i),P_i=p_i)$ is not physically 
equivalent to the 
system defined by the \h \ (\ref{z}) and deformed commutation relations (\ref{3}), 
although they have a  common energy spectrum and a common 
dynamical symmetry group. None the less,
all physical quantities in a noncommutative system can be determined in terms of 
relevant quantities in a transformed system, expressed in terms of cannonical 
variables.

\section{Conclusion}

The importance of a consistent Fock-space picture in physics is 
well understood, but we wish to stress that this picture is especially suited
for the analysis of  many-body problems and for the discussion of  symmetries 
in the problem. Also, the construction of the ladder operator is an 
interesting mathematical problem connected with (quantum) group 
theory \cite{borzov}, and  has its applications in physical 
chemistry \cite{chem}.

Any eigenvalue problem $H\Psi_n=E_n\Psi_n$ with  a discrete
spectrum bounded from below, can be described in  Fock space as 
$\{ b^{\dagger n}|0\rangle; n=0,1,2,\ldots\}$ with $\Psi_n= b^{\dagger n}
|0\rangle/n!;\;[b,b^{\dagger}]=1$.
So far, there has been no  simple method of expressing 
$b^{\dagger}=b^{\dagger}(x,p)$, that would
generalize  the  simple relation $b^{\dagger}=(x-ip)/\sqrt{2}$  valid 
for the ordinary \ho .
A general approach to the construction of  generalized ladder operators
exists only  for a certain class
of exactly solvable potentials, namely, the   shape-invariant potentials
\cite{bal}.
The solution of this problem is connected with the recursive
relations among orthogonal polynomials $P_n$, defined as
$\Psi_n=\Psi_0 P_n$.

We  have explicitly constructed deformed ladder operators using the exact solutions
found in Ref.\cite{Chang} and have discussed the corresponding deformed
Fock space.  
These results  represent a step towards  finding a general method 
for constructing ladder operators for any smooth deformation of cannonical 
commutation relations.
The dynamical symmetry for the 
described problem is  a group of rotations $O(D)$, and 
is defined by a specific choice of deformed commutation relations Eq.(\ref{3}).
A slight generalization that we have suggested 
offers  new possibilities for symmetry breaking.

Finally, let us note that the case $\beta+\beta'<0$ but $D\beta+\beta'>0$ was
not discussed in the literature. This case might be physically interesting 
since it indicates that there is singularity in the system (transformation
Eq.(\ref{4}) becomes singular) when ${\bf p}^2_{\rm crit}=-1/(\beta+\beta')$.
This value  has to be very large, ${\bf p}^2_{\rm crit}\lsim
E_{\rm Pl}$, so it implies that relativistic effects have to be considered 
(nonrelativistic quantum mechanics is not applicable anymore!). 
This singularity might suggest the generalization of special relativity
with new (invariant) scale, connected with the Planck scale, and 
similar to the generalizations proposed in Refs.\cite{ac}. The modified
relativistic quantum mechanics approach to the minimal length
uncertainty relations is very important in order to get consistent  physical
picture.

Acknowledgments

We thank M. Milekovi\'c for useful discussions.
This work was supported by the Ministry of Science and Technology of the
Republic of Croatia under contracts No. 0098002 and No. 0098003.

\end{document}